\documentclass{aa}  

\usepackage{graphicx}
\usepackage{txfonts}
\usepackage{hyperref}
\hypersetup{
    colorlinks=true,
    linkcolor=cyan,
    citecolor=cyan,
    urlcolor=cyan,
    }

\newcommand{\orcid}[1]{\protect\href{https://orcid.org/#1}{\protect\includegraphics[width=8pt]{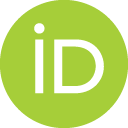}}}

\begin{document}

   \title{Revisiting symbiotic binaries with interferometry}
   
    \subtitle{II. New PIONIER data}
    
   \author{Henri M. J. Boffin\orcid{0000-0002-9486-4840}\thanks{Based on ESO observations done under Prog. ID 114.2714.}
        \inst{1}
          \and
          Jaroslav Merc\orcid{0000-0001-6355-2468}\inst{2,3}
          }

    \authorrunning{Henri M. J. Boffin and Jaroslav Merc}

   \institute{European Southern Observatory, Karl-Schwarzschild-Stra\ss{}e 2, Garching bei M\"{u}nchen, 85748, Germany
         \and
             Astronomical Institute of Charles University, V Hole\v{s}ovi\v{c}k{\'a}ch 2, Prague, 18000, Czech Republic
             \and
             Instituto de Astrof\'isica de Canarias, Calle Vía Láctea, s/n, E-38205 La Laguna, Tenerife, Spain
             }

   \date{Received June 27, 2025; accepted August 1, 2025}

% \abstract{}{}{}{}{} 
% 5 {} token are mandatory
 
  \abstract
   {Symbiotic stars, which generally comprise a red giant and an accreting white dwarf, are excellent laboratories to understand mass transfer in wide binaries, with application to a wide family of systems. One of the fundamental questions is how mass is transferred from the red giant to the white dwarf. We use interferometric measurements made with the VLTI/PIONIER instrument, combined with {\it Gaia} data, to measure the radius of the giant in seven symbiotic systems. We further place the giants in the H-R diagramme, which allows us to estimate their mass and to show that they are all very evolved and likely on the asymptotic giant branch. We compare our measured giant radii to their Roche-lobe radius and show that, except for ZZ CMi, all giants are well within their Roche lobe and that mass transfer likely takes place via stellar wind. Our interferometric data provide further evidence that the giant in ZZ CMi (nearly) fills its Roche lobe. Our conclusions are still hampered by the poor characterisation of some of the giants or their binary orbit, and we encourage the community to make an effort to provide these.  
   }


   \maketitle
%
%-------------------------------------------------------------------

\section{Introduction}

Symbiotic stars are among the most enigmatic binary systems in astrophysics, comprising a compact accretor -- typically a white dwarf -- interacting with an evolved, mass-losing red giant in a binary system with orbital period between hundreds of days and several years \citep{1982ASSL...95.....F,2009syst.book.....K,2019arXiv190901389M,2025Galax..13...49M}. This interplay drives complex photometric and spectroscopic variability, fueled by accretion processes, thermonuclear shell burning, and occasional nova-like outbursts \citep{2013A&A...559A...6L,2025CoSka..55c..47M}. 

Their importance spans multiple astrophysical domains: they serve as laboratories for studying accretion physics in detached binaries, the late stages of stellar evolution, and potential progenitors of Type Ia supernovae. Moreover, their outflows contribute to the chemical enrichment of the interstellar medium, making them key players in galactic ecology. Despite their significance, their formation pathways, outburst mechanisms, and evolutionary fates remain open questions.

An interesting question is how mass is transferred in these systems \citep{2025ApJ...980..224V}.  \cite{1999A&AS..137..473M} found that symbiotics are, almost without exception, detached binaries, in which case the mass transfer took place via stellar wind \citep{2015ASSL..413..153B}. This is likely not so surprising if we assume typical masses of 1.5~M$_\odot$ and 0.5~M$_\odot$ for the giant and white dwarf, respectively, and thus a mass ratio of 3. In this case, according to the most recent canonical binary evolution models, we would expect that if the giant fills its Roche lobe, the mass transfer would be dynamically unstable, and quickly a common envelope would form, leading to a very close binary with an orbital period of a few days or less \citep{2020cee..book.....I}. Symbiotic systems wouldn't thus exist! Things are more complicated, however, as in some cases, the white dwarf appears more massive than the giant, and Roche lobe filling is fully possible \citep[e.g., T CrB; ][]{2025A&A...694A..85P,2025ApJ...983...76H}. Symbiotic systems are, by nature, known to be accreting, and this seems difficult to reconcile with the low efficiency expected from stellar wind accretion, especially if the giants are not very evolved \citep[see, however, the recent work by][]{2025arXiv250211325M}. 

Therefore, it is critical to determine the evolutionary stage of red giants in symbiotic systems and the fraction of the Roche lobe they are filling. 

\defcitealias{2025A&A...695A..61M}{Paper~I}
Interferometry is here most useful as it allows to determine the angular diameter of the giant, $\Phi$. Combined with the knowledge of the orbital parameters and of the distance of the system, we can then obtain the filling fraction. This was already done by \cite{2014A&A...564A...1B} and then revisited and extended by \citet[][hereafter \citetalias{2025A&A...695A..61M}]{2025A&A...695A..61M}, who used PIONIER archival data. Here, we analyse a new set of acquired PIONIER interferometric data for seven additional symbiotic stars and derive the filling fraction in them, confirming that all but one of them do not appear to be filling their Roche lobe. 
%--------------------------------------------------------------------

\begin{table*}[hptp]%
\centering
\caption{VLTI/PIONIER observation log of our targets and the resulting uniform disc (UD) angular diameters, with the errors estimated from bootstrapping. We also indicate when the first and last science observations on a given night were obtained. When data were obtained on more than one night, we also compute the UD for the combined data.\label{tab:log}}%
%\tabcolsep=0pt%
%\begin{tabular}{20pc}{@{\extracolsep\fill}lcc@{\extracolsep\fill}}
\begin{tabular}{lccccc}
\hline\hline
\noalign{\smallskip}
Object & Configuration & First Science & Last Science & UD & $\chi^2$ \\
 &    &Observation & Observation &   (mas)      & \\
% &    &            &             &  & \\
 \hline
 \noalign{\smallskip}
V1044 Cen	&	Extended&		2025-01-02T07:53:50	&	2025-01-02T08:12:34&0.550 $\pm$ 0.005		&0.35\\
			&			&2025-01-07T07:27:23	&	2025-01-07T07:43:02&0.557 $\pm$ 0.011		&0.63\\
            &           &  & combined& 	0.552 $\pm$ 0.005		&0.50\\		
\noalign{\smallskip}
V417 CMa	&	Extended	&	2024-11-28T03:34:52	&	2024-11-28T04:30:36& 0.589 $\pm$ 0.005	&	0.79\\
\noalign{\smallskip}
SY Mus		& Extended	&	2025-01-08T05:07:48	&	2025-01-08T06:47:53&0.719 $\pm$ 0.005		&0.46\\
\noalign{\smallskip}
V420 Hya	&	Extended	&	2025-01-04T07:18:07		&2025-01-04T07:33:10&0.864 $\pm$ 0.011			&1.81\\
\noalign{\smallskip}
V648 Car	&	Large	&		2024-12-24T06:01:28		&2024-12-24T06:18:55&0.906 $\pm$ 0.010	&1.06\\
			&			&2024-12-26T04:01:35	&	2024-12-26T04:16:38&0.915 $\pm$ 0.006		&0.87\\
            &           &  & combined& 0.910 $\pm$ 0.005		&0.97\\
\noalign{\smallskip}
Hen 3-461	&	Large	&		2024-12-24T06:47:18	&	2024-12-24T07:03:38&1.100 $\pm$ 0.004		&0.43\\
\noalign{\smallskip}
ZZ CMi		& Large		&	2024-12-23T07:15:23	&	2024-12-23T08:19:33&1.642 $\pm$ 0.009 &	3.82\\
\noalign{\smallskip}

\hline
\noalign{\smallskip}
%\hline
\end{tabular}
\end{table*}

\section{Observations and data analysis}
Observations of seven symbiotic systems were done over the period from 27 November 2024 to 8 January 2025 (see Tab.~\ref{tab:log}) with the PIONIER instrument \citep{2011A&A...535A..67L} at the Very Large Telescope Interferometer in Paranal, Chile, under Prog. ID 114.2714 (PI: Merc). The 1.8-m Auxiliary Telescopes were used in the large (3 targets) or extended (4 targets) configurations, with baselines ranging between 49 and 202 metres. Observations were made in Service Mode, and for each target, two sequences of 45 minutes (CAL-SCI-CAL-SCI-CAL) were obtained -- some of these taken continuously, some on different nights -- for a total of 10.5 hours. 
The choice of the large over the extended configuration was based on the estimated diameter of the giant star in the symbiotic system. 
The raw data are available in the ESO archive\footnote{\url{archive.eso.org}} and were processed with the {\tt pndrs} pipeline \citep{2011A&A...535A..67L}. 

\begin{figure*}[h]
   \centering
   \includegraphics[width=0.99\textwidth]{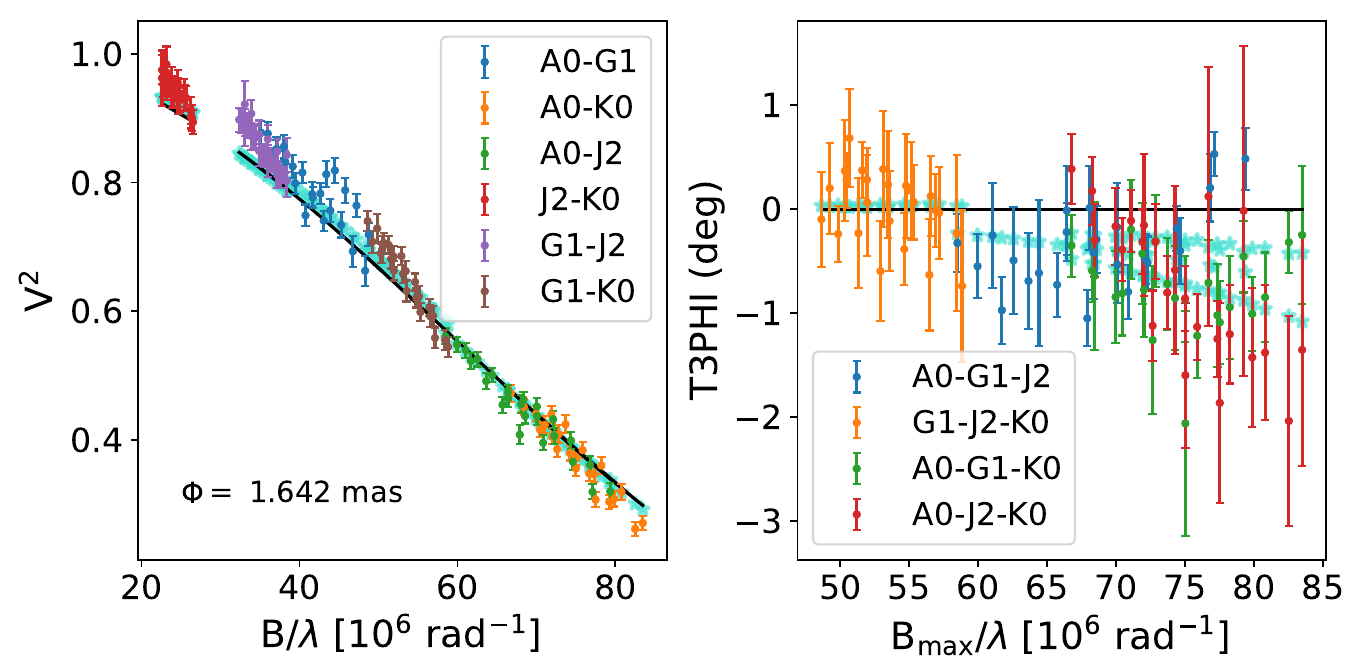}
      \caption{Reduced PIONIER interferometric data for ZZ CMi. The left plot shows the squared visibilities ($V^2$), while the right one shows the closure phases  (T3PHI) in degrees. The colours correspond to the various baselines ($B$, expressed in terms of the wavelengths, $\lambda$), while the black line is the simple fit of a uniform diameter. The turquoise stars are the fit corresponding to a Roche-lobe filling (and, hence, deformed) star. They are mostly different from the black line in the closure phases.  
              }
         \label{fig:ZZCMi}
   \end{figure*}

\begin{figure*}[h]
   \centering
   \includegraphics[width=0.95\textwidth]{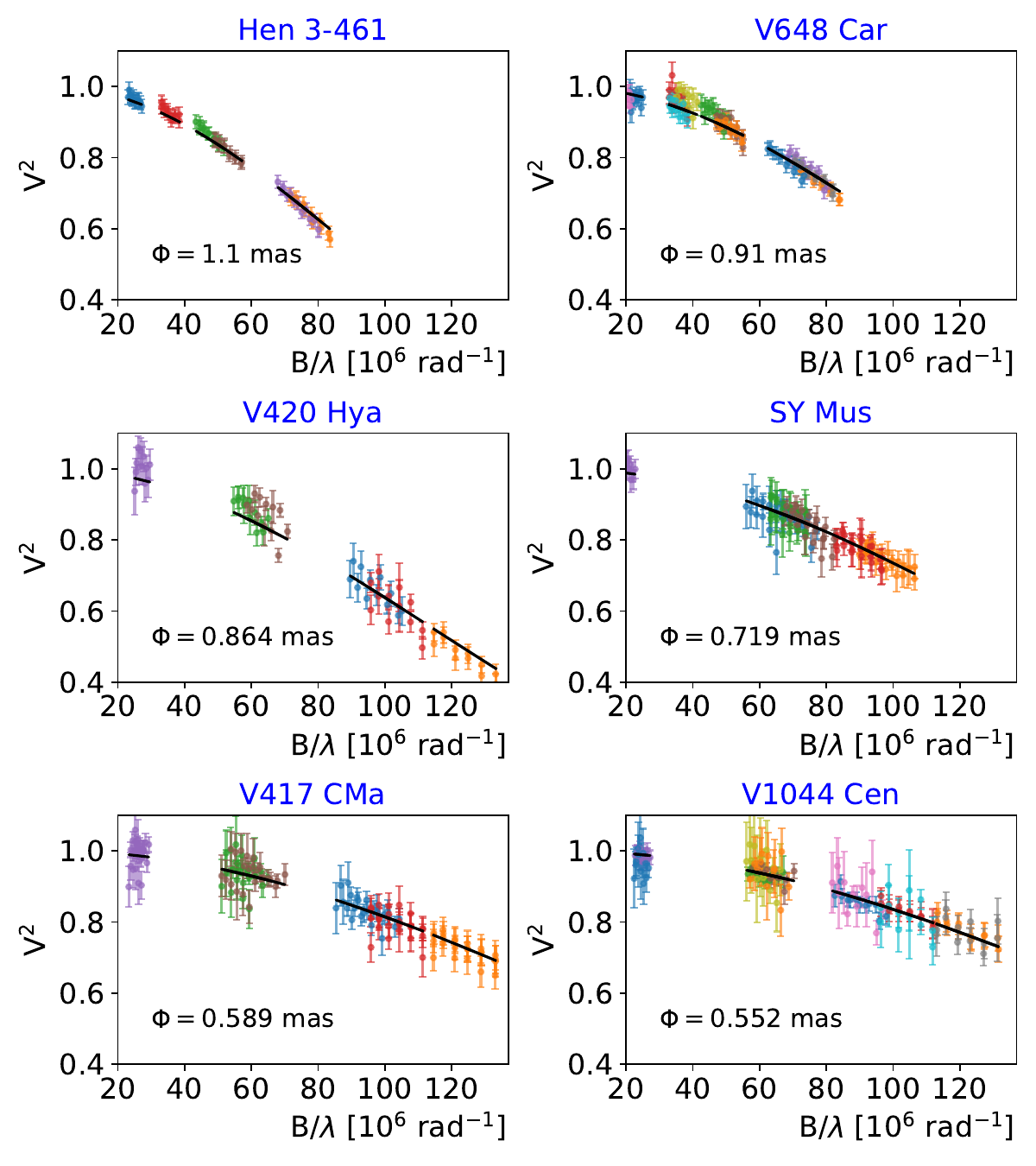}
      \caption{Squared visibilities for the remaining six symbiotic systems as a function of the baselines. The colours correspond to the various baselines, while the black line is the fit. The angular diameter that is then obtained is also indicated.
              }
         \label{fig:V2}
   \end{figure*}

We use {\tt PMOIRED}\footnote{\url{https://github.com/amerand/PMOIRED}} \citep{2022SPIE12183E..1NM} to analyse the reduced interferometric data. In all but one case (see the worst case example in the right panel of Fig.~\ref{fig:ZZCMi}), the closure phases did not show any significant deviations from zero and were thus not considered further. We fit a uniform disc, using only the squared visibilities, and use a bootstrapping with 8192 fits to estimate the error bars. We have checked that we obtained the same value for the angular diameter using LITPro\footnote{The LITpro software is available at \url{http://www.jmmc.fr/litpro}} \citep{2008SPIE.7013E..44T}, as well as our own script. 
In \citetalias{2025A&A...695A..61M} of this series, we have shown that there is almost no difference between the derived diameter for a uniform disc and a limb darkened one, with the second one depending on the choice of the limb darkening model and its parameters. As the main error on the resulting stellar radius is due to the (much larger) relative error on the parallax, using the model-free uniform disc appears better suited. Our results, together with the resulting chi-squared, are shown in Tab.~\ref{tab:log} (where we ordered the observations according to the measured angular radius) as well as in Figs.~\ref{fig:ZZCMi} and \ref{fig:V2}. The latter clearly indicate that, except perhaps for ZZ CMi, a uniform disc is a good fit to the data. We note, as in \citetalias{2025A&A...695A..61M}, that for several systems, the $\chi^2$
value is smaller than 1, likely indicating that the errors on the squared visibilities computed by the pipeline are overestimated. The properties of the red giant and the orbital elements of our systems are indicated in Tab.~\ref{tab:target_stars}. 

\section{Targets}

 \begin{figure*}
   \centering
   \includegraphics[width=0.99\textwidth]{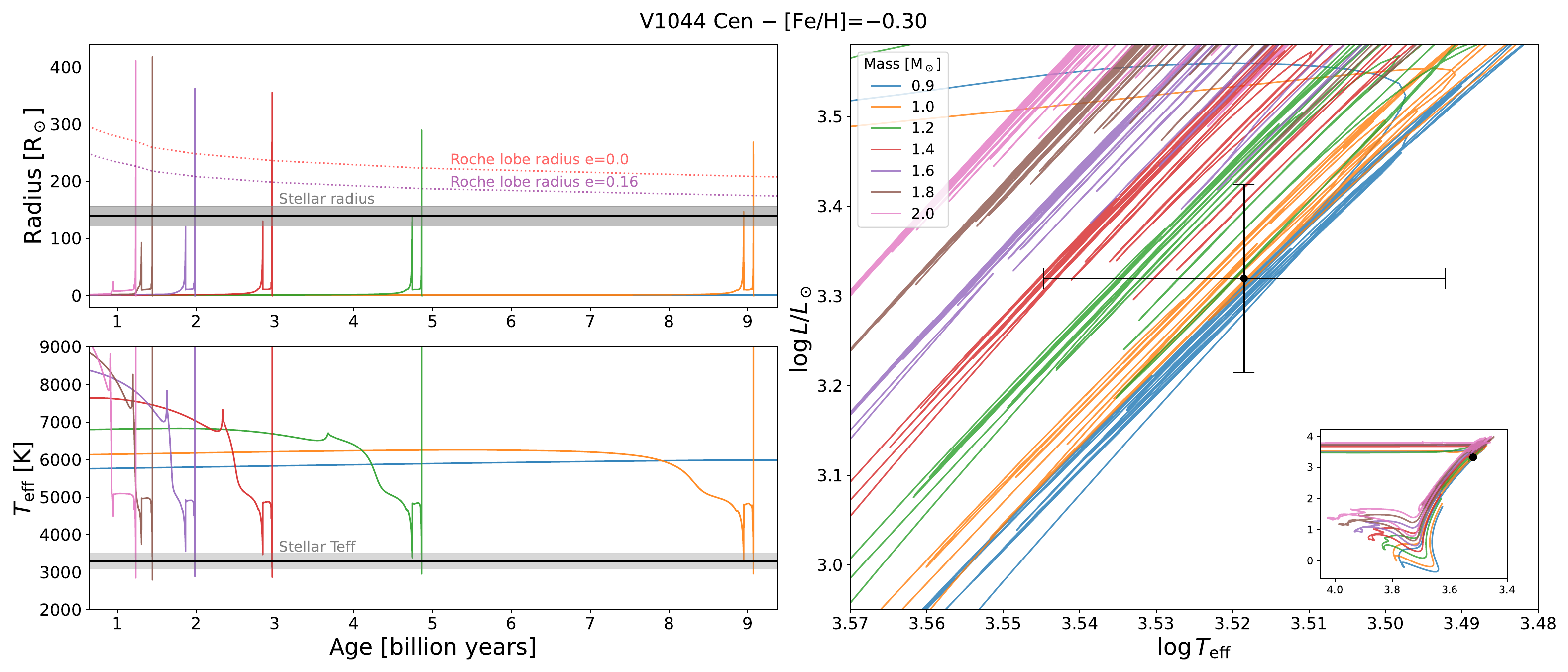}
      \caption{Comparison between our derived properties and MIST stellar models for V1044 Cen for various masses. The  left panels show the radius (upper) and effective temperature (lower) as a function of age, for various stellar masses, with the derived quantities indicated with the black lines and the gray shaded regions. For the radius, we also show the Roche lobe radius for a circular orbit (red dotted line) and at periastron for $e=0.16$ (purple dotted line). The right panel shows the H-R diagramme, with the inset showing the whole region. Our derived quantities are indicated with the black point and the error bars.
              }
         \label{fig:HR_V1044}
   \end{figure*}

\begin{table*}%
\centering
\caption{Effective temperatures and metallicities of the cool components of the target stars, along with their orbital elements. For references, please see the text. \label{tab:target_stars}}%
\begin{tabular}{lllllll}
\hline\hline
\noalign{\smallskip}
Object & T$_{\rm eff}$ & [Fe/H] & $P$ & $e$ & $f(m)$ \\
 &  (K) & (dex) &    (days) &  & (M$_\odot$)\\
 \noalign{\smallskip}
\hline
\noalign{\smallskip}
V1044 Cen & 3300 & $-0.3$ & 985.0 $\pm$ 5.4 & 0.158 $\pm$ 0.057 & 0.0083$\pm$0.0014 \\
V417 CMa & 3550 & 0.0: &  397.5 $\pm$ 0.1 & 0.042 $\pm$ 0.002 & 0.0198 $\pm$ 0.0001 \\
SY Mus & 3400 & $-0.15\pm0.08$ & 624.36 $\pm$ 0.80 & 0.0  & 0.0302 $\pm$ 0.0017 \\
V420 Hya & 3400  & 0.0 & 751.4 $\pm$ 0.2 & 0.099 $\pm$ 0.004  & 0.089 $\pm$ 0.001  \\
V648 Car & 3500: & ... &  560: & ... &  ... \\
Hen 3-461& 3200 & 0.12 $\pm$ 0.11 &  2271 $\pm$ 18 & 0.404 $\pm$ 0.045 & 0.086 $\pm$ 0.013  \\
ZZ CMi & 3250: & ...& 983:&... &... \\
\hline
\end{tabular}
\end{table*}

\subsection{V1044 Cen}\label{sec:v1044cen}
The most distant symbiotic star from our sample, and thus not surprisingly, also the one with the smallest angular diameter (0.552 $\pm$ 0.005 mas), V1044 Cen (Hen 3-886; CD $-36^{\circ}8436$), hosts an M5.5 giant. 
\cite{2016MNRAS.455.1282G}
derive  an effective temperature T$_{\rm eff}$ of 3\,300~K, a metallicity  ([Fe/H]) of $-0.30$, and a projected rotational velocity, $v \sin i_* = 8.1~\pm~1.1$~km/s,  where $v$ is the stellar rotation velocity and $i_*$ is the inclination of the rotation axis of the star on the plane of the sky.

\textit{Gaia} DR3 gives a parallax of 0.425 $\pm$ 0.051 mas \citep{2016A&A...595A...1G,2023A&A...674A...1G}, implying a distance of $2352^{+327}_{-254}$~pc and, based on our measurements, a radius for the giant of $140^{+19}_{-15}$~R$_\odot$ and a luminosity of $2087^{+563}_{-438}$~L$_\odot$. The RUWE\footnote{RUWE stands for the Renormalised Unit Weight Error associated to each \textit{Gaia} source.
The RUWE is expected to be around 1.0 for sources where the single-star model provides a good fit to the astrometric observations.} value of 1.359 is, however, scarily above the 1.25 threshold expected for a safe astrometric solution \citep{2022MNRAS.513.2437P}. 

In their series of papers on symbiotic system orbits, \cite{2015AJ....150...48F} found an orbital period ($P$) of 985.0 $\pm$ 5.4 d, an eccentricity ($e$) of 0.158 $\pm$ 0.057, and a spectroscopic mass function ($f(m)$) of 0.0083 $\pm$ 0.0014~M$_\odot$. Despite the small eccentricity and its large error, these authors still propose that the eccentric solution should be preferred over the circular one -- although this must be marginally so. The difference in the reduced chi-square is indeed only 12\% between the two solutions. If we take the eccentricity at face value, then one should consider the pseudo-synchronisation period of 853.4 d, according to the above authors. 

At face values, both synchronisation with the orbital period or the pseudo-synchronisation period wouldn't be possible, as they require a value of $\sin i_* > 1.$ When taking all errors into account, the inclination can, however, span a wide range between 30 and 90 degrees. With 30 degrees and the pseudo-synchronisation period, we obtain $v \sin i_* = 4.2 $~km/s, almost half the nominal observed value. However, in giants, there can be other broadening mechanisms that could easily reconcile these values. Moreover, there is no reason to assume that synchronisation occurred in a star whose radius is continuously changing. 

Using the effective temperature and luminosity, we can place the object in an Hertzsprung-Russell (H-R) diagramme and, compared with stellar evolutionary models, determine the mass of the giant, $m_g$. Given the low metallicity, the MIST models\footnote{\url{https://waps.cfa.harvard.edu/MIST/interp_tracks.html}}  \citep{2016ApJS..222....8D} and the parameters indicated above lead to a mass of the giant of $1.0^{+0.4}_{-0.1}$~M$_\odot$ (Fig.~\ref{fig:HR_V1044}), with masses below 0.9~M$_\odot$ being impossible given that they would imply a star older than the Universe. For $m_g \leq 1.2$~M$_\odot$, it is not possible to affirm if the giant is still on the tip of the first red giant branch (RGB) or already on the asymptotic giant branch (AGB). In any case, as shown in the upper left panel of the figure, once on the AGB, it will fill its Roche lobe after a few thermal pulses. 

A direct consequence of the low mass of the giant, is that the companion will also be rather low mass: $0.27^{+0.06}_{-0.02}$~M$_\odot$ (assuming an inclination of the orbital plane on the sky, $i$, of 60$^\circ$). This would imply the companion to be an He WD and that the evolution of its progenitor got interrupted quite early on the RGB -- a very improbable scenario. Indeed, as the current giant is now already more evolved, this would imply that the orbital period {\it increased} following the first mass transfer episode. 
Only if the orbital inclination would be as small as $\approx 30^\circ$ would the companion mass be again within the range of values compatible with being a CO WD. It would indeed be $0.54^{+0.11}_{-0.03}$~M$_\odot$ ($i$ = 30$^\circ$).
In this case, the Roche lobe is $209^{+27}_{-8}$~R$_\odot$ and the filling factor is thus $0.67^{+0.11}_{-0.15}$, or $0.80^{+0.13}_{-0.18}$ at periastron in case of an eccentric orbit. 

 \begin{figure*}
   \centering
   \includegraphics[width=0.99\textwidth]{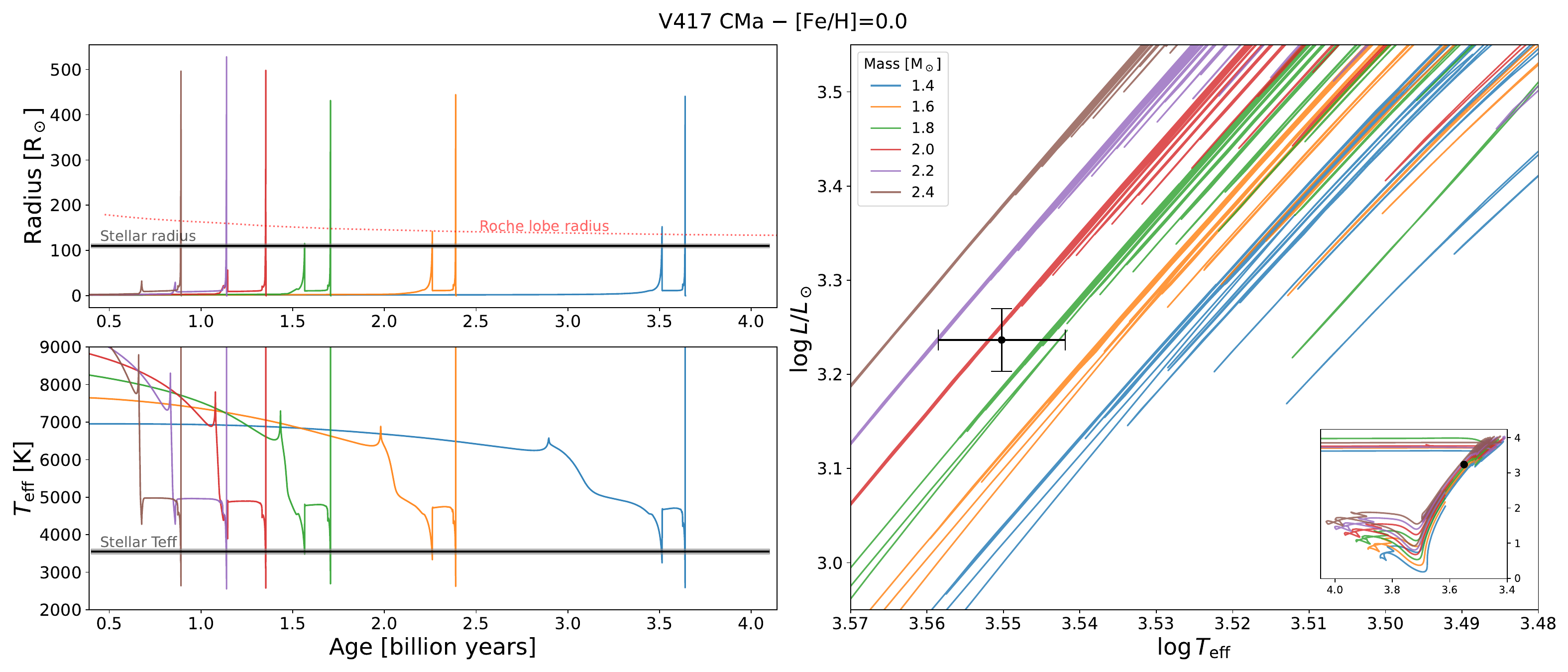}
      \caption{Same as Fig.~\ref{fig:HR_V1044} for V417 CMa, assuming a solar metallicity.
              }
         \label{fig:HR_V417}
   \end{figure*}

    \begin{figure}
   \centering
   \includegraphics[width=0.49\textwidth]{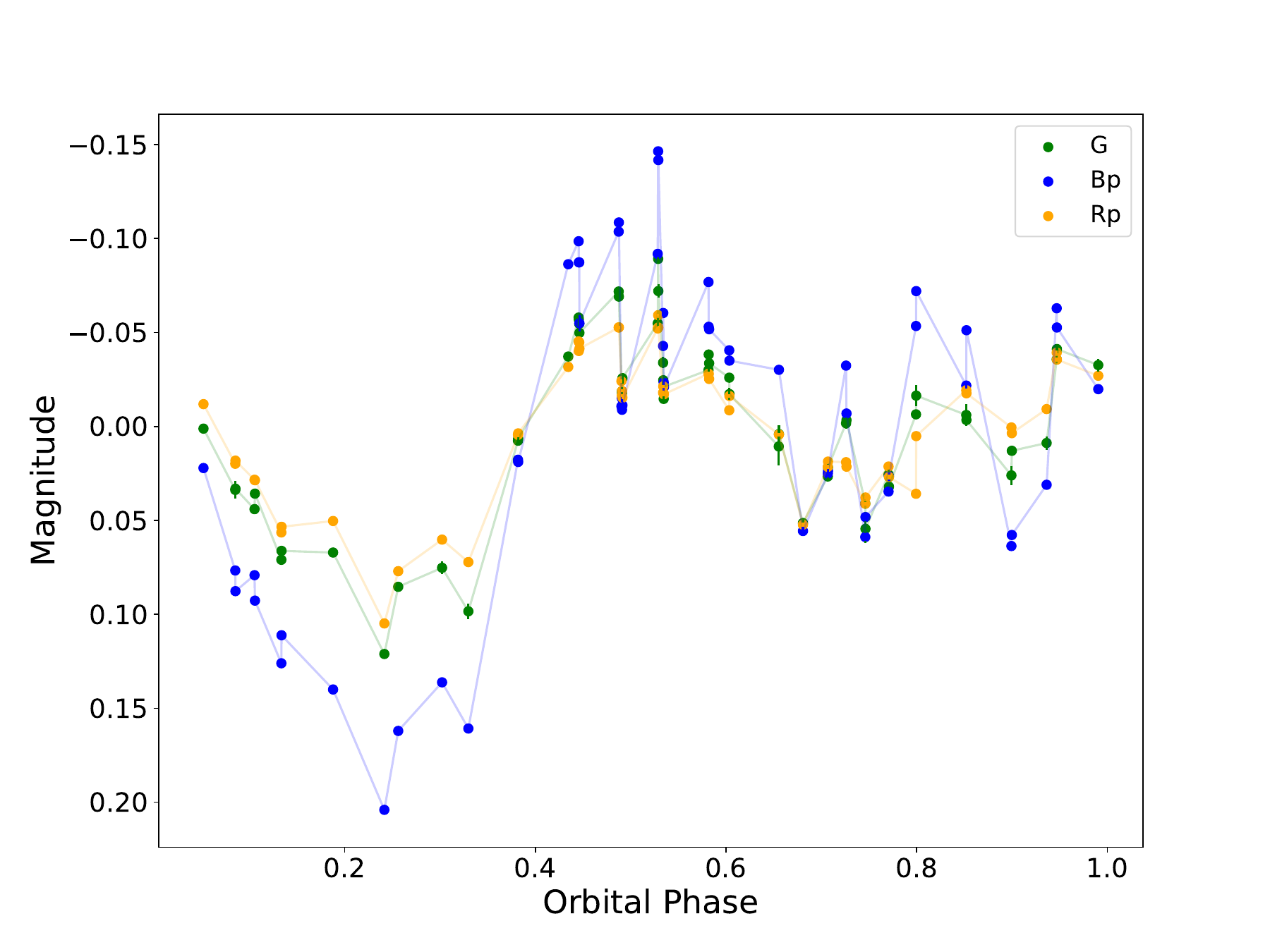}
      \caption{The {\it Gaia} light curve of V417 CMa (in the three bands, $G$, $B_{\rm P}$, $R_{\rm P}$), phase folded on a period of 397.5 days.
              }
         \label{fig:HR_V417_Gaia}
   \end{figure}
   
\subsection{V417 CMa}\label{sec:v417cma}
This symbiotic star -- also known as CD$-$28$^\circ$3719 and \mbox{Hen 3-21} -- comprises an extrinsic S-type star, that is, an S star lacking Tc and whose s-process enhancement therefore likely originates from mass transfer from a companion \citep{2002A&A...396..599V}. 
\cite{2019A&A...626A.127J} derived a preliminary orbit with a period of 397.5 days and an eccentricity of 0.042 $\pm$ 0.002. The residuals of the orbit are about ten times larger than the individual errors on the radial velocities, which indicates the likely presence of an additional source of noise, for example pulsations. This would also explain the non-zero eccentricity in this case. 
It presents $\delta$-type X-ray emission, typical of symbiotic stars that are powered by accretion \citep{2010ATel.3053....1L,2024A&A...689A..86L}. \citet{2024A&A...683A..84M} reported detection of flickering using the \textit{TESS} data.

\textit{Gaia} DR3 gives a parallax of 0.575 $\pm$ 0.022 mas, leading to a distance of 1\,739 $\pm$ 67 pc, although the RUWE of 1.269 is just above the safe limit to completely trust the solution. 
\cite{2000A&AS..145...51V} derive an apparent bolometric magnitude of 7.85. Given the \textit{Gaia} distance, this translates to an absolute bolometric magnitude of $-3.35$ -- similar to what is found by \cite{2022A&A...664A..45A} -- and a luminosity of 1724 L$_\odot$. 

We derive an angular diameter of 0.589 mas -- surprisingly close to the parallax, which translates to a linear radius of 110 $\pm$ 5 R$_\odot$. With the above luminosity, this corresponds to an effective temperature of 3\,550 $\pm$ 68 K, in the expected range for such a S star. This is in agreement with what is found by \cite{2017A&A...601A..10V}, who also quote a metallicity of $-0.5$, albeit with little confidence.  

 The mass of the giant that we can infer from the MIST stellar evolutionary models depends crucially on the metallicity, which is thus not well constrained. For [Fe/H]=0, we find (Fig.~\ref{fig:HR_V417}) $m_g=1.95 \pm 0.25$~M$_\odot$; for [Fe/H]=$-0.25$, $m_g=1.4 \pm 0.2$~M$_\odot$; and for [Fe/H]=$-0.5$, $m_g=1.0 \pm 0.2$~M$_\odot$. 
Because we see a clear signal (of about 0.15 mag amplitude in \textit{Gaia} $B_{\rm P}$) in the \textit{Gaia} light curve with the same period as the orbital period  (Fig.~\ref{fig:HR_V417_Gaia}), the orbital inclination, $i$, cannot be lower than about 70 degrees. Moreover, as the giant star presents s-process overabundance that must come from pollution from the companion, the companion must have gone through enough thermal pulses to be itself polluted before finishing its life as a white dwarf. Thus, the current companion must have a mass, $m_c$, of at least 0.5-0.6~M$_\odot$, depending on the metallicity and the initial progenitor’s mass — the lower the metallicity, the higher $m_c$. Given the orbit’s spectroscopic mass function, $f(m) = 0.0198 \pm 0.0001$, this imposes a minimum value of the giant current mass: for $m_c=0.5$ M$_\odot$ and the most favourable case of $i=70\deg$, we have $m_g \geq  1.8$ M$_\odot$, while for $i=80\deg$, this becomes 1.96 M$_\odot$. The corresponding values for $m_c=0.55$ M$_\odot$ are 2.08 M$_\odot$ and 2.28 M$_\odot$, and for $m_c=0.6$ M$_\odot$, 2.41 M$_\odot$ and 2.63 M$_\odot$. Obviously, the mass of the progenitor of the white dwarf had to be more massive than the {\it original} mass of the current giant (and not much more to avoid a common-envelope evolution, which would have led to a much shorter orbit than the current one), but this is not very constraining as we do not know how much mass the latter accreted (and it had to accrete to be polluted in s-process elements). One can thus assume that the current giant mass is about the same as the companion’s progenitor mass. 

Another point to consider is that low-mass stars reach a larger radius on the first giant branch (Fig.~\ref{fig:HR_V417}) and thus, given the orbital period, a star less massive than about 1.6~M$_\odot$ will likely fill its Roche lobe while on the first giant branch. As the current giant is an S star, likely at the start of the AGB, this cannot have happened. Thus, such a mass is also a minimum mass for the giant. 
All this seems to converge to the conclusion that the star must have almost solar metallicity in order to have a consistent solution and be at least 1.8 M$_\odot$. One can thus take as representative values those corresponding to solar metallicity: $m_g = 2$~M$_\odot$ and $m_c \approx 0.54$ M$_\odot$. In that case, the Roche lobe radius of the giant is 153~R$_\odot$ and the filling factor is then 72\%. The giant is likely just on the AGB and it will fill its Roche lobe as soon as it undergoes its first thermal pulse. 

\begin{figure*}[ht]
   \centering
   \includegraphics[width=0.99\textwidth]{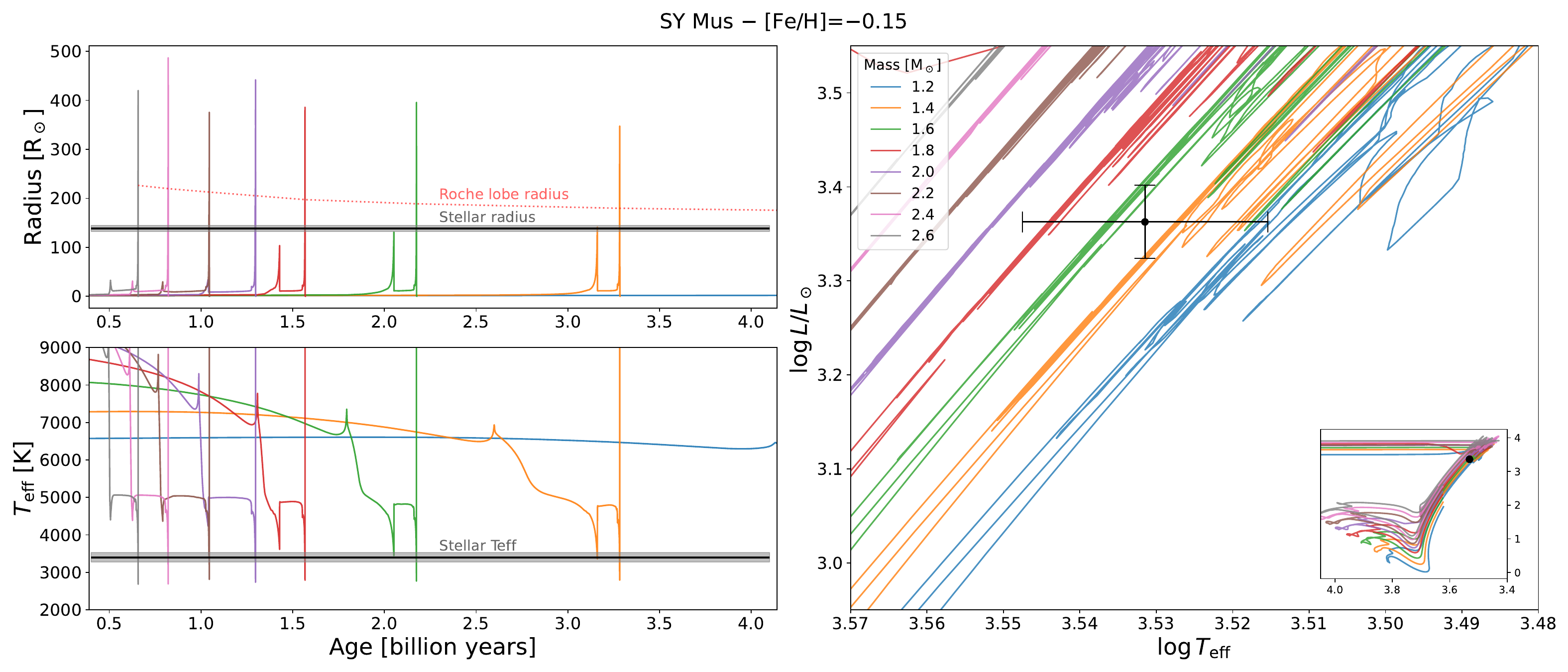}
      \caption{Same as Fig.~\ref{fig:HR_V1044} for SY Mus.
              }
         \label{fig:SYMUs}
   \end{figure*}

\subsection{SY Mus}\label{sec:symus}
This system contains an M5 giant, with an estimated effective temperature of 3\,400~K and an almost solar metallicity of [Fe/H] = $-0.15 \pm 0.08$ \citep{2016MNRAS.455.1282G}, for which \textit{Gaia} provides a parallax $\varpi = 0.558 \pm 0.025$. Quite remarkably, the RUWE for this system is 1.063, and would thus indicate that the astrometric solution can be trusted. The parallax implies a distance of 1794$^{+85}_{-78}$ pc, which combined with our angular diameter of 0.719 mas, leads to a radius of $138.6^{+6.6}_{-6.1}$~R$_\odot$ and thus a luminosity of  $2305^{+224}_{-199}$~L$_\odot$. 
Our derived value for the radius is very similar to the one estimated by \cite{2007BaltA..16...49R} -- 135~R$_\odot$ -- from the analysis of the light curve.  These authors also reported the presence of ellipsoidal variability. 

\cite{2017AJ....153...35F} found the orbit of SY Mus to be circular, with an orbital period of 624.4 d and a spectroscopic mass function, $f(m)=0.0302$. They further use an inclination of 84 degrees, in line with several previous works (see their Sec. 6.1) that report that the system is eclipsing. By comparing with MIST evolutionary tracks (Fig.~\ref{fig:SYMUs}), we estimate that the giant has a mass of $1.5 \pm0.4$~M$_\odot$. Assuming $i=84^{\circ}$ leads then to a companion mass of $0.50 \pm 0.07$~M$_\odot$. The Roche lobe radius is then about $184 \pm 20$~R$_\odot$ and the filling factor, $75\pm9$\%. If the giant has a mass on the lower side of the permitted range, then it could still be on the tip of the RGB. In most cases, it will be on the AGB, however. It is interesting to note that despite the low filling factor, the system apparently presents ellipsoidal variability -- this was already the case for the systems V1261 Ori and RW Hya that were analysed in \citetalias{2025A&A...695A..61M}. 

\cite{2016MNRAS.455.1282G} determined a rotation velocity, $v \sin i_* =6.6\pm0.6$ km/s, which when synchronisation is assumed, would imply an inclination of 36 degrees, too small to explain the light curve. With $i = 84^{\circ}$, we have to assume, however, that the rotation period is much larger than the orbital period, being at least equal to 1075 days.

 \begin{figure*}
   \centering
   \includegraphics[width=0.99\textwidth]{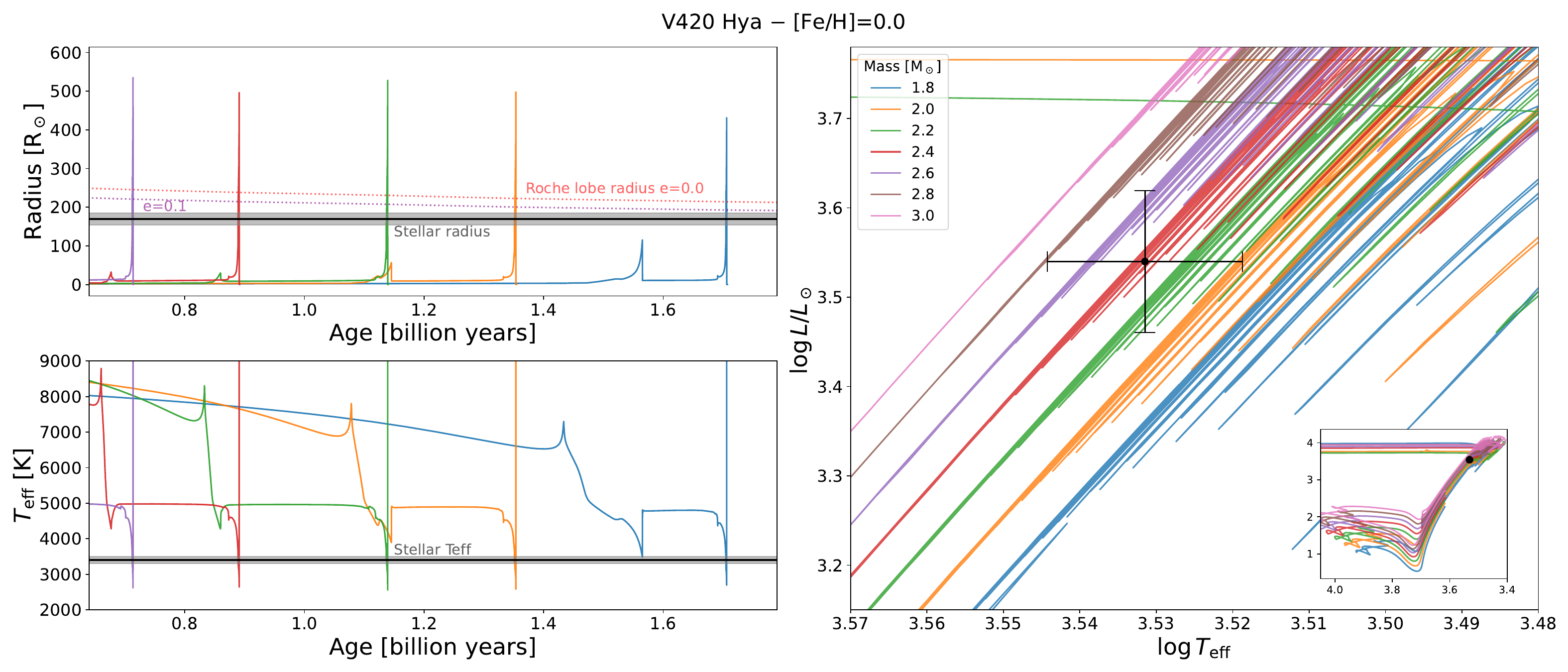}
      \caption{Same as Fig.~\ref{fig:HR_V1044} for V420 Hya.
              }
         \label{fig:V420Hya}
   \end{figure*}
   
\subsection{V420 Hya}\label{sec:v20hya}
This object, also known as CD$-27^\circ8661$, Hen 3-785, or Hen 4-121, is like V417 CMa, an extrinsic S star, with the giant having the spectral type S4. It is, in fact, one of the most evolved extrinsic S stars \citep{2002A&A...396..599V}. \textit{Gaia} provides a parallax of $0.547 \pm 0.050$ mas, leading to a distance of $1824^{+186}_{-149}$ pc, although the RUWE of 1.388 is unsatisfactorily high. \cite{2012BaltA..21...39J} derive an effective temperature of 3\,400 K and a solar metallicity. Combined with our angular diameter estimate of 0.848 mas, this implies a stellar radius of $166^{+17}_{-14}$~R$_\odot$ and a luminosity of $3306^{+677}_{-557}$~L$_\odot$. The comparison with MIST stellar evolutionary tracks (Fig. \ref{fig:V420Hya}) leads to a giant mass of $2.4\pm0.4$~M$_\odot$. Based on the radius and effective temperature, the giant is likely on the AGB, which is in line with the spectral classification. 

\cite{2012BaltA..21...39J} find the H$\alpha$ line profile to vary as a function of the orbital phase, while \textit{Gaia} reveals photometric variations with an amplitude of 0.11 mag. This hints to a high inclination for this system. In fact, \cite{2002A&A...396..599V} report an eclipse of the ultraviolet Balmer continuum by the S star, while \cite{2013AN....334..860A} included it in their catalogue of eclipsing binaries.  The system is also classified as a $\delta$-type symbiotic star, based on X-rays \citep{2024A&A...689A..86L}. Similarly as for V417 CMa, \citet{2024A&A...683A..84M} reported detection of flickering.

\cite{2019A&A...626A.127J} derive an eccentric orbit ($P=751.4\pm0.2$~d, $e=0.099\pm0.004$; $f(m)=0.089\pm0.001$~M$_\odot$), with however, some large residuals, likely indicative of pulsations. Assuming an inclination of 90 degrees, this leads to a rather massive companion ($1.01^{+0.06}_{-0.10}$ ~M$_\odot$), and a Roche lobe radius of $238^{+14}_{-16}$~R$_\odot$. Taking into account the eccentricity, this implies at periastron, a Roche-lobe filling around $79\pm5$\%. If the inclination is smaller, then the companion is more massive and the Roche-lobe filling fraction slightly larger.  

\begin{figure*}
   \centering
   \includegraphics[width=0.99\textwidth]{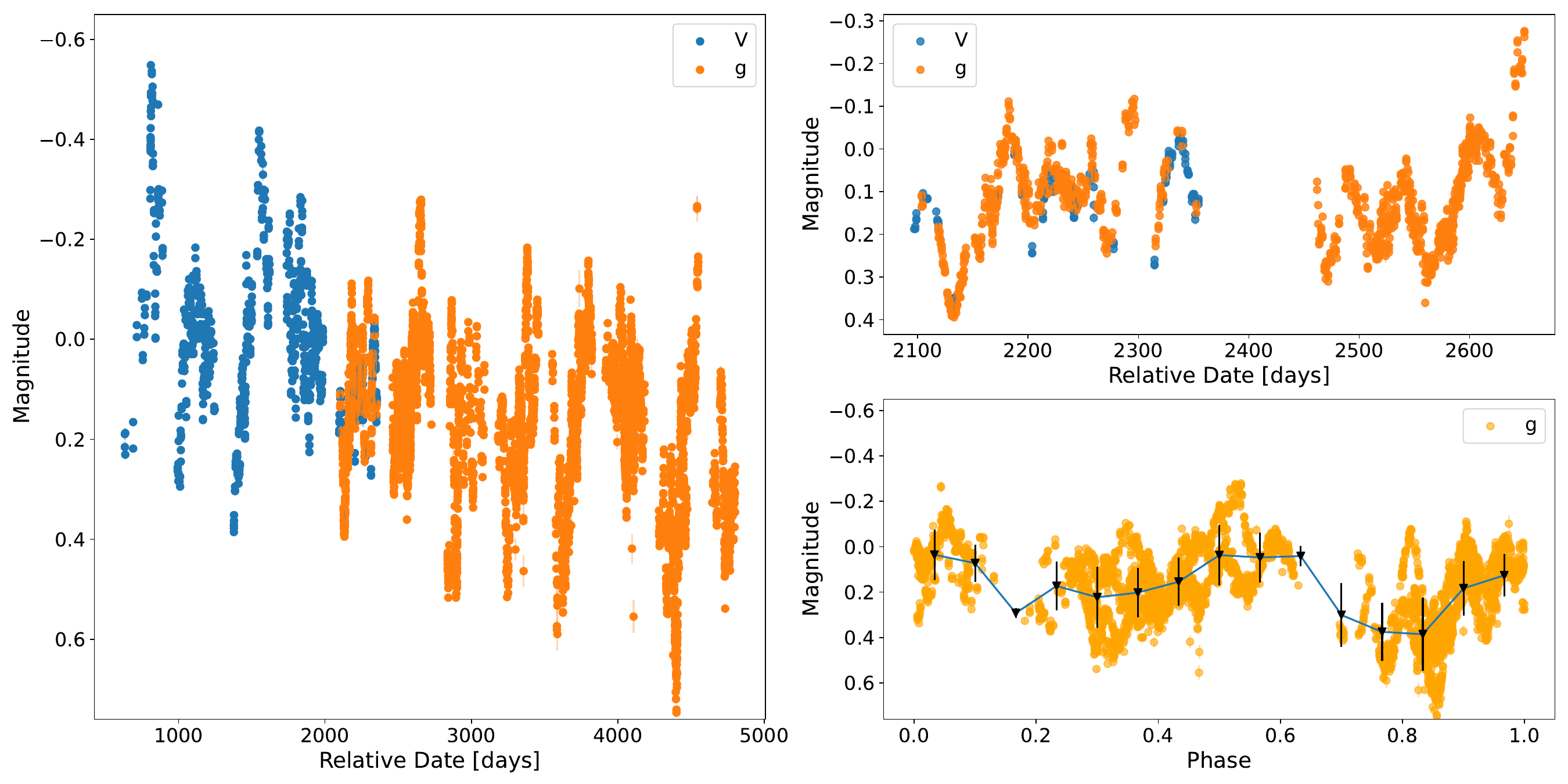}
     \caption{\label{fig:V420_lc} The ASAS-SN light curve of V420 Hya is shown on the left, with magnitudes in the $V$- and $g$-band. The $g$-band observations were shifted by 0.18 mag to bring them in line with the $V$ ones. The upper right panel shows a zoom-in region of 600 days to highlight the short-term periodicity. The lower right panel shows the $g$-band data folded on the best period we determined, 767.364 days, together with some phase-averaged mean to illustrate the apparent ellipsoidal or eclipsing variations.
              }
  \end{figure*}
   
We downloaded\footnote{\url{https://asas-sn.osu.edu/sky-patrol/}} the All-Sky Automated Survey for SuperNovae \citep[ASAS-SN;][]{2014ApJ...788...48S,2019MNRAS.485..961J} light curve, which is shown in Fig.~\ref{fig:V420_lc}. A periodogramme reveals a peak at a period of 383.682~d, which we take as half the orbital period (767.364~d, close to the spectroscopically derived one). The data phased on this latter value indeed show an unequal double peaked light curve, typical of ellipsoidal variations. The presence of this signal also implies that the inclination must be rather high, while the Roche-lobe filling fraction seems rather small to explain the light curve. At the shorter periods, the periodogramme also has a peak at 57.5~d, which is likely the pulsation period of the giant. The light curve indeed shows high-amplitude pulsations, in line with the large residuals of the spectroscopic orbit. 

  \begin{figure*}
   \centering
   \includegraphics[width=0.99\textwidth]{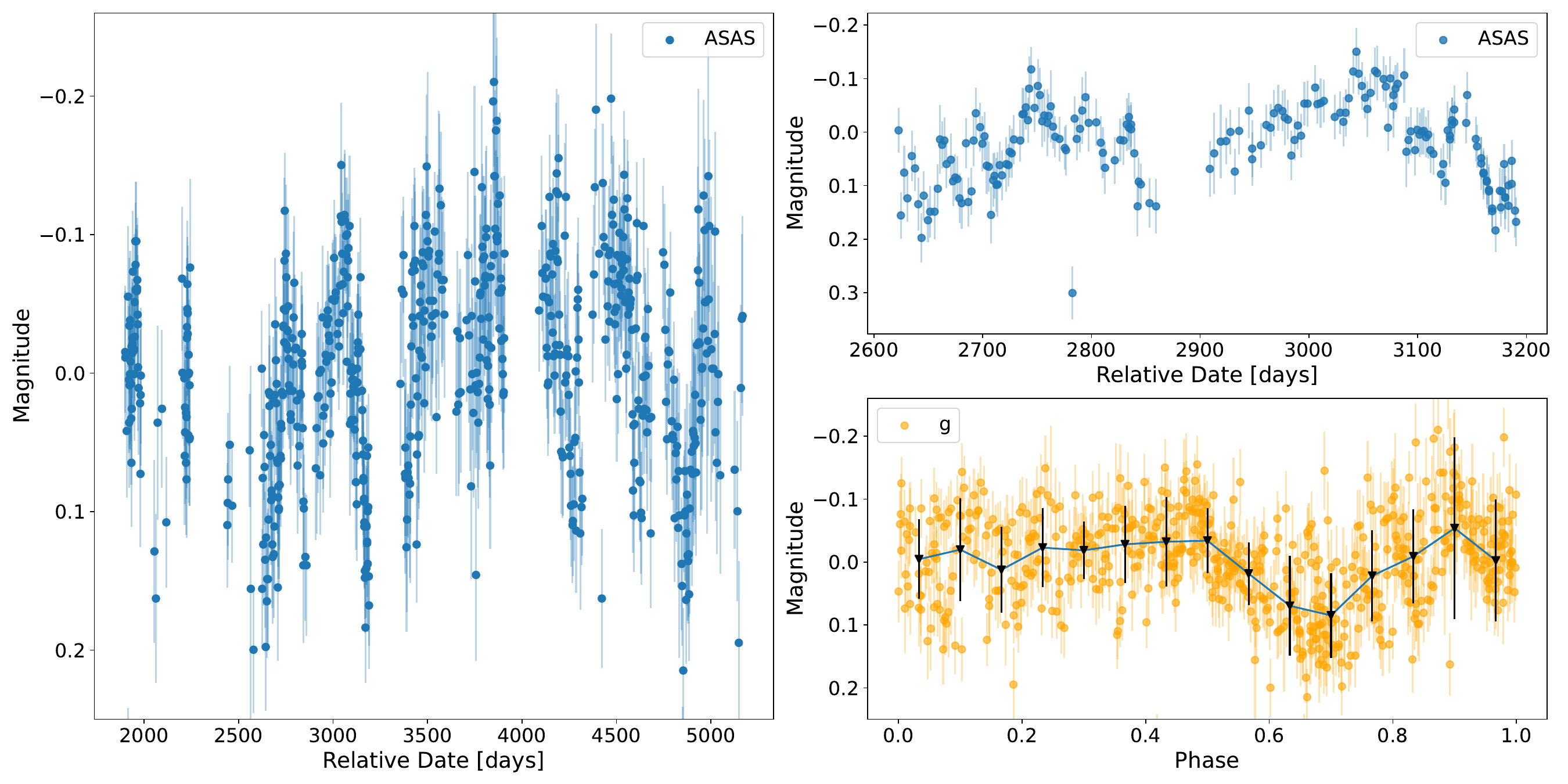}
     \caption{\label{fig:v648_lc} Same as Fig.~\ref{fig:V420_lc} for V648 Car. The lower right panel presents the data when folded on the period of 560.39 days.}      
   \end{figure*}

\subsection{V648 Car}\label{sec:v648car}
This system contains an M4 III giant with a clear blue excess \citep{2003A&A...397..927P}. It belongs to the class of hard X-ray emitting symbiotics \citep{2024A&A...689A..86L}, whose origin is likely the boundary layer between the accretion disc and the white dwarf \citep{2010ApJ...709..816E}. 
\cite{2012ApJ...756L..21A} found the system to be flickering with very large amplitude -- later confirmed by \cite{2024A&A...683A..84M} using \textit{TESS} observations and detected as well in X-rays \citep[see refs in][]{2024A&A...689A..86L}, further indicating that it is an accreting symbiotic star. \cite{2012ApJ...756L..21A} also determined from the light curve the presence of a tidally distorted giant and a binary period of $\approx520\pm68$ days. 

The ASAS catalogue of variable stars\footnote{\url{https://www.astrouw.edu.pl/asas/?page=acvs}} quotes on the other hand a periodicity of 577 days \citep{2003AcA....53..341P}. Our re-analysis of the same set of data (Fig.~\ref{fig:v648_lc}) indicates a periodicity at 280.1958~days, which we will tentatively assume is half the orbital period -- which is then 560.39~d. The light curve shows possibly an ellipsoidal variability, although it could also more be an eclipse, superimposed on pulsations with much shorter periods. Pending more precise data, it is premature to state that this is a case similar to SY Mus and V420 Hya. 

 \begin{figure*}
   \centering
   \includegraphics[width=0.99\textwidth]{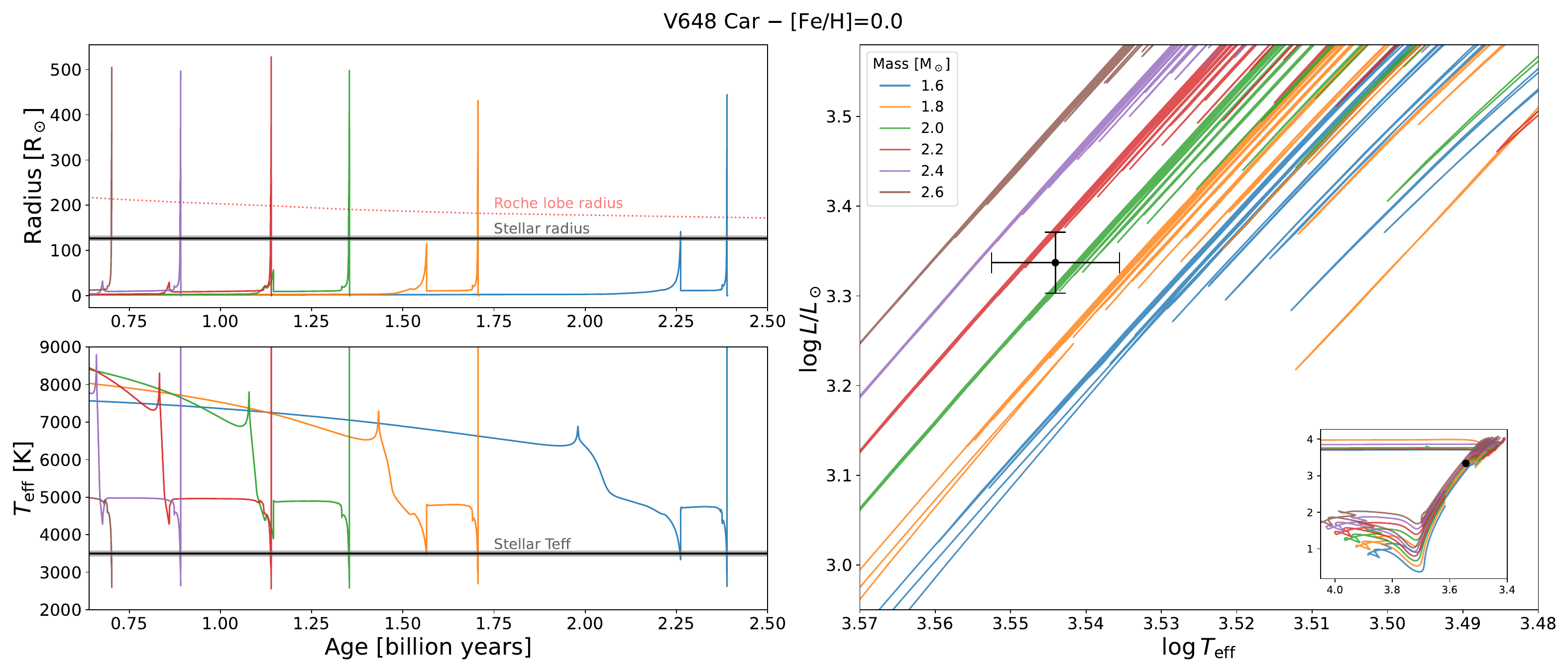}
      \caption{Same as Fig.~\ref{fig:HR_V1044} for V648 Car, assuming solar metallicity.
              }
         \label{fig:V648Car}
   \end{figure*}
   
\textit{Gaia} quotes a parallax of $0.7727\pm0.0302$ mas and, hence, a distance of $1294^{+54}_{-48}$~pc, which may need to be taken with some caution given the relatively large RUWE of 1.355. 
We derive an angular diameter of 0.91 mas, which translates to a linear radius of $ 127 \pm 5$~R$_\odot$. 
The calibration of \cite{1999AJ....117..521V} allows us to transform the M4 spectral type to an effective temperature of $\approx$3\,500~K. When putting the star in a H-R diagramme (Fig.~\ref{fig:V648Car}), assuming solar metallicity, we derive a mass of the giant of 2.1 $\pm$ 0.3~M$_\odot$, while this value becomes $1.4^{+0.2}_{-0.1}$~M$_\odot$ if the metallicity is $-0.3$. Only for the smaller masses, can the giant still be on the RGB -- for the others, it is on the AGB. 

If we assume that the orbital period is indeed close to 560 days, and a canonical 0.6~M$_\odot$ mass for the potential white dwarf companion, then a mass of  1.4~M$_\odot$ for the giant leads to a Roche lobe radius about 163 $\pm$ 10 R$_\odot$, and thus a filling fraction of 0.78 $\pm$ 0.05. For a giant of 2.1~M$_\odot$, the Roche lobe radius is 195 $\pm$ 12 R$_\odot$, and the filling fraction is even smaller. The Roche lobe radius would be smaller by about 10~R$_\odot$ if the companion was a 1~M$_\odot$ star instead. In all cases the filling fraction is smaller than about 0.85. 

 \begin{figure*}
   \centering
   \includegraphics[width=0.99\textwidth]{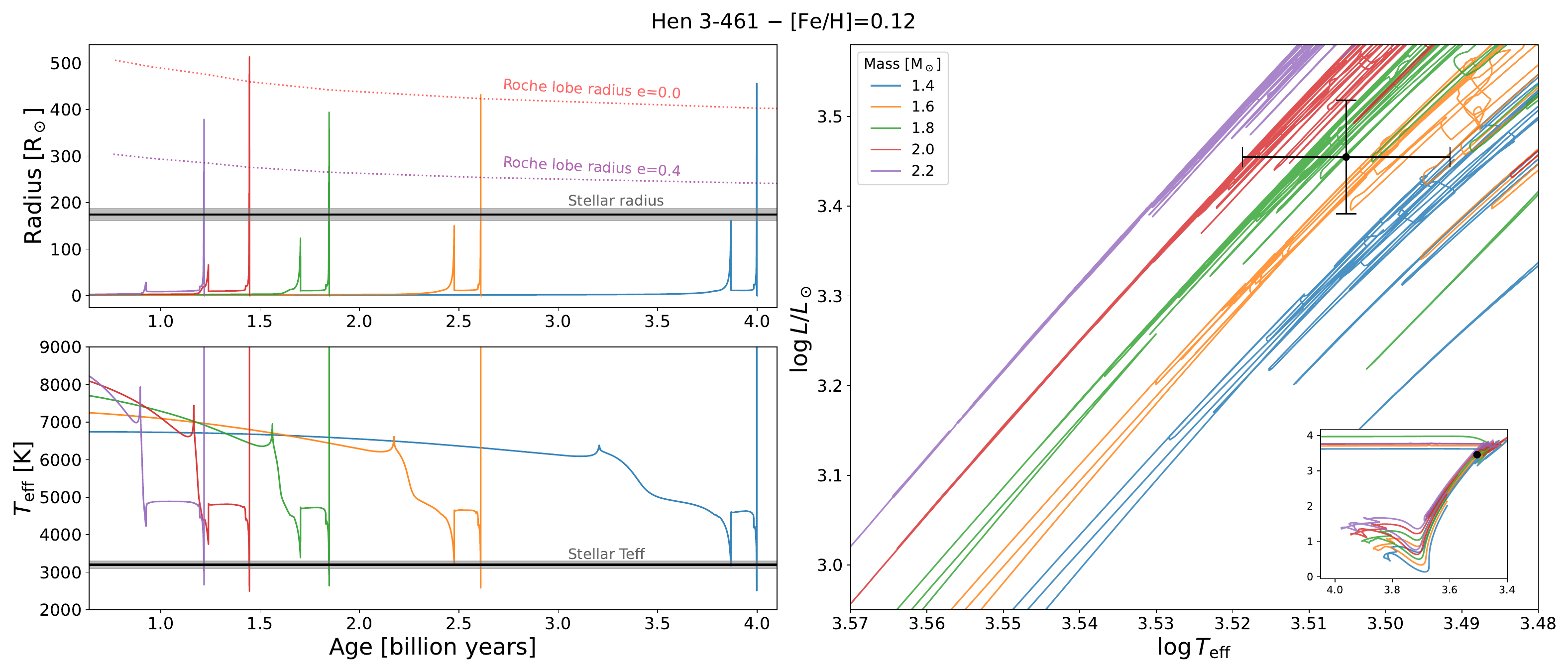}
      \caption{Same as Fig.~\ref{fig:HR_V1044} for Hen3-461.
              }
         \label{fig:Hen3}
   \end{figure*}
   
\subsection{Hen 3-461}\label{sec:hen3461}
The giant in this system is classified as M7, making it one of the coolest symbiotic systems known \citep{1999A&AS..137..473M}, with \cite{2016MNRAS.455.1282G} 
quoting a value for T$_{\rm eff}$ = 3200 $\pm$ 100 K, a solar metallicity (0.12 $\pm$ 0.11), and a $v \sin i_*= 7.4 \pm 0.6$~km/s. X-ray emission from the system was detected by \citet{2016ApJ...824...23N}, who classified it as $\beta$/$\delta$-type, indicating contributions from both the boundary layer of the accretion disc and shock-heated plasma in the wind collision region. More recently, \citet{2024A&A...683A..84M} reported flickering in the optical light curve, consistent with the presence of an accretion disc.

\textit{Gaia} DR3 provides a parallax of 0.678 $\pm$ 0.049 mas for Hen 3-461, which translates to a distance of $1474^{+115}_{-99}$~pc, with a relatively safe RUWE of 1.21. \textit{Gaia} also reported photometric variations, but without any period, while based on ASAS data, \cite{2013AcA....63..405G} report a period of 635 days. This is much less than the orbital period, 2\,271 $\pm$ 18 d, reported by \cite{2017AJ....153...35F}, who also indicate a rather eccentric orbit, with $e = 0.404 \pm 0.045$ and a spectroscopic mass function of $f(m)=0.086 \pm 0.013$~M$_\odot$. 

We derive an angular diameter of $1.10 \pm 0.01$ mas, leading to a radius of $174\pm14$~R$_\odot$. With this value, and assuming pseudo-synchronisation corresponding to the 0.4 eccentricity, that is, a period of 1056 d \citep{2017AJ....153...35F}, leads to an inclination $i = 62 \pm 10$ degrees. The radius and the effective temperature lead to a luminosity of 2850 $\pm$ 815 L$_\odot$, and comparison with the MIST stellar tracks (Fig.~\ref{fig:Hen3}), imply a mass of the giant of $1.7\pm0.3$~M$_\odot$. Given the mass function, this means a rather massive companion of $0.79^{+0.11}_{-0.06}$~M$_\odot$ ($i=90^{\circ}$) or $0.93^{+0.12}_{-0.07}$~M$_\odot$ ($i=62^{\circ}$). If the inclination is even smaller than 62 degrees, then the companion would be even more massive. 

The giant can be on the tip of the RGB only if it has among the smallest possible masses. For most of the masses, it is more likely on the AGB. Even with such a long period, the star will still fill its Roche lobe before the end of its evolution. However, for the moment, it only fills about 40\% of its formal Roche lobe, which is equivalent to 66\% of the Roche lobe at periastron, given the high eccentricity. 

 \begin{figure*}
   \centering
   \includegraphics[width=0.99\textwidth]{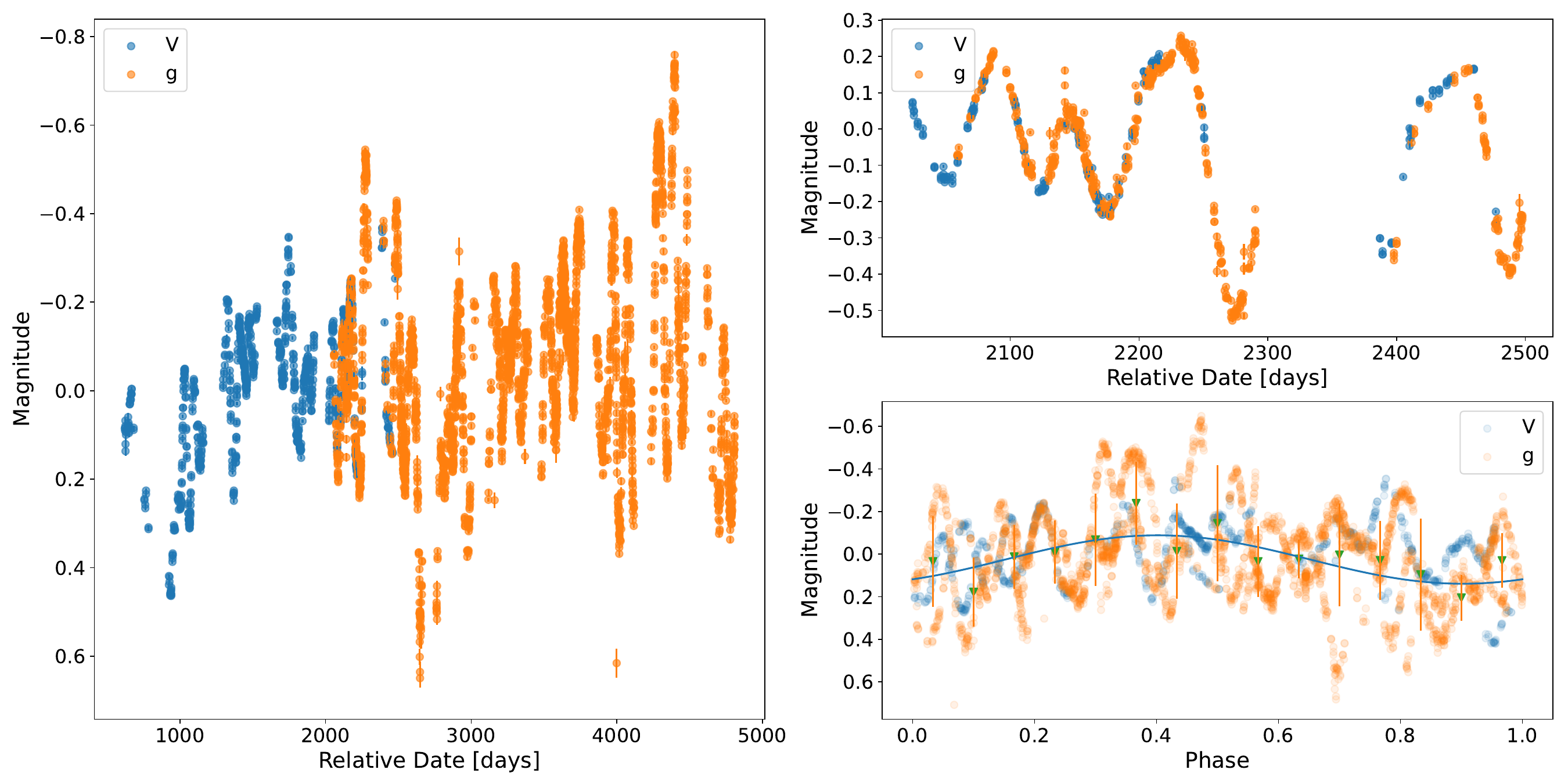}
     \caption{\label{fig:ZZCmi_lc} Same as Fig.~\ref{fig:V420_lc} for ZZ CMi. The $g$-band magnitudes were shifted by $-0.06$ mag to bring them in line with the $V$-band ones. The lower right present the data when folded on the period of 983 days.}      
   \end{figure*}

 \begin{figure*}
   \centering
   \includegraphics[width=0.99\textwidth]{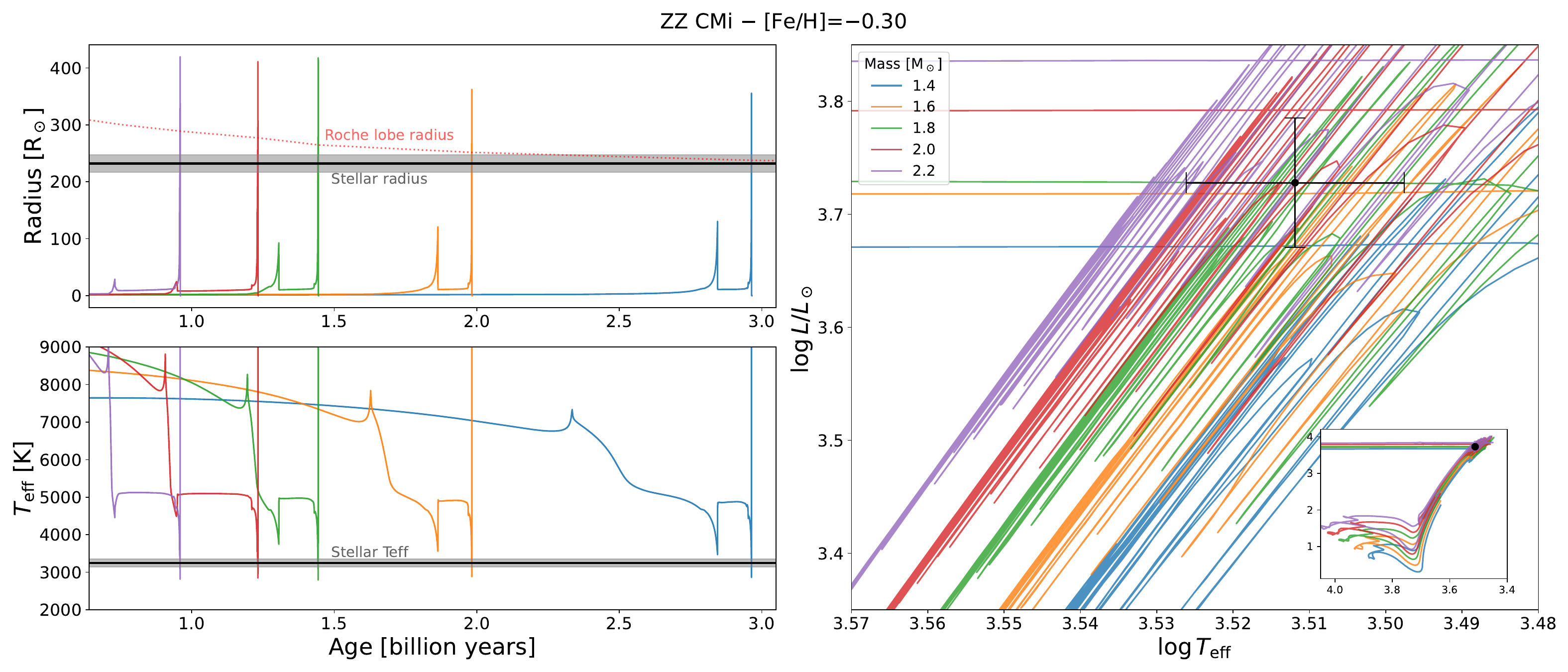}
      \caption{Same as Fig.~\ref{fig:HR_V1044} for ZZ CMi, assuming [Fe/H]=$-0.30$.
              }
         \label{fig:ZZCMi-hr}
   \end{figure*}

\subsection{ZZ CMi}\label{sec:zzcmi}
The giant in this system has a CDS SIMBAD spectral type of M6 I-IIep, which would normally imply that it is a supergiant and thus a massive star. If that were correct, this would not be a symbiotic star in the canonical sense.
However, \cite{2021AN....342..952Z} infer that ZZ CMi contains an M4-M6 III giant with an accreting white dwarf, a classification confirmed by \cite{2024SerAJ.208...41Z}. It is also known to be characterised by two X-ray thermal components, both soft
and hard, that is sharing features of $\beta$ and $\delta$ types. The soft
emission is most likely produced in a colliding-wind region \citep{1997A&A...319..201M}, and the hard emission is most likely
produced in an accretion-disc boundary layer \citep{2013A&A...559A...6L}. Flickering was also reported by \cite{2024A&A...683A..84M}, further indicating that accretion is taking place.

\cite{2013AcA....63..405G} quote a period of 106 days, which they attribute to pulsations, while using also photometry, \cite{2010arXiv1003.0608W} claim an orbital period of 437.18 d, with the observed signal being due to an ellipsoidal effect, although the light curves do not appear very clean. 

The star was observed by ASAS-SN \citep{2014ApJ...788...48S,2019MNRAS.485..961J} over a period of 4\,000 days in the $V$-band (3\,085 measurements) and, later, in the $g$-band (870  measurements), with some overlap. We show the light curve in Fig.~\ref{fig:ZZCmi_lc}, which clearly reveals a short-term periodicity. To combine both sets of observations, we apply a correction of $-0.06$ mag to the $g$-band measurements. A periodogramme analysis then provides a period of 112.31906 days, which is likely the pulsation or rotation period of the giant. The semi-amplitude of this signal varies between 0.1 and 0.4 mag. A periodogramme on values binned over the above derived period indicates a new period of 982.98 days, which is possibly the orbital period. The relevant signal is only $0.11 \pm 0.03$ mag, smaller than the main signal, but still marginally significant at 3.66 $\sigma$. The residual do not provide any signal when phased with a period of 437.18 days. \cite{2010arXiv1003.0608W} do mention that the much nicer U-band light curve shows a peak in the periodogramme at 883.9 days, closer to the orbital period we infer, of 983 days. 
The periodogramme on the ASAS-SN raw data indicates also a second period of 105.95413 days, similar to what was found by \cite{2013AcA....63..405G}.

\begin{table*}%
\centering
\caption{Representative inferred parameters for our targets -- for the full range and uncertainties, please see text. \label{tab:results}}%
\begin{tabular}{lllccc}
\hline\hline
\noalign{\smallskip}
Object & Radius & Luminosity & Mass & Mass & Roche-lobe  \\
 & & & giant & companion & filling fraction\\
 & (R$_\odot$) & (L$_\odot)$ &    (M$_\odot)$ &  (M$_\odot$) & (\%)\\
 \noalign{\smallskip}
\hline
\noalign{\smallskip}
V1044 Cen & 140 & 2087 & 1.0 & 0.54 & 0.67-0.80\\
V417 CMa & 110 & 1724 &  2.0 & 0.54 & 0.72 \\
SY Mus & 139 & 2305 & 1.5 & 0.5 & 0.75   \\
V420 Hya & 166  & 3306 & 2.4 & 1  & 0.79 \\
V648 Car & 127 & 2173 &  1.4-2.1 & ... & $<0.85$\\
Hen 3-461& 174 & 2850 &  1.7 & 0.8-0.9 &  0.66\\
ZZ CMi & 231 & 5350 & 2.4& 1: & $>0.9$\\
\hline
\end{tabular}
\end{table*}

The \textit{Gaia} DR3 parallax value is 0.76 $\pm$ 0.05 mas, leading to a distance of $1315.5^{+92.5}_{-80.6}$ pc -- a value that should be safe both due to the small relative error and the small value of the RUWE, 1.015. 
We derive a uniform angular disc diameter of 1.641 mas, which, when coupled with the parallax, leads to a physical radius of 231 $\pm$ 16~R$_\odot$. 

Using the calibration of \cite{1999AJ....117..521V}, we can associate the M6 spectral type to an effective temperature of $\approx$3\,250~K. The luminosity we then infer is 5\,350 $\pm$ 750~L$_\odot$, making ZZ CMi, the most evolved of all our targets, as also indicated by its spectral luminosity class. Given its evolutionary status, the star must be on the AGB and one should have undergone some thermal pulses. It should therefore show s-process enrichment and a spectrum would be useful to confirm this.  

From the H-R diagramme (Fig.~\ref{fig:ZZCMi-hr}), we find that for [Fe/H]=$-0.3$, the giant has a mass of $1.8\pm0.4$~M$_\odot$, while for solar metallicity, this would be $2.4\pm0.4$~M$_\odot$. Assuming a 0.6~M$_\odot$ companion, we find a Roche-lobe radius of $264\pm27$~R$_\odot$ for [Fe/H]=$-0.3$ and $300\pm23$~R$_\odot$ for [Fe/H]=0.0, leading to a Roche-lobe filling fraction of, respectively, 0.88 $\pm$ 0.09 and 0.77 $\pm$ 0.08. If the companion is a 1.0~M$_\odot$ object, the Roche lobes decrease to $252\pm22$~R$_\odot$ and $285\pm19$~R$_\odot$, and the filling fractions increase proportionally, to 0.92 and 0.81. 

Our interferometric measurements indicate the possibility that ZZ CMi is close to filling its Roche lobe. Indeed, ZZ CMi is the object for which the $\chi^2$ value is the largest (Table~\ref{tab:log}), indicating a particularly poor fit of the visibilities with a simple disc. We have therefore followed the procedure of \cite{2024A&A...692A.218M} to compute the visibilities and the closure phases corresponding to a Roche lobe filling star. The projected shape of such a star depends on the mass ratio, orbital inclination, and orbital phase, being most elongated at or near quadrature. However, since the mass ratio, inclination, and orbital elements (specifically, the time of quadrature) are not known for ZZ CMi, and since changes in inclination and orbital phase produce similar effects on the projected shape, we assumed an inclination of 90\textdegree{} and generated models for three orbital phases (exact quadrature, and phases rotated by 15\textdegree{} and 30\textdegree{}), and for five mass ratios ranging from 1.2 to 3.6. During fitting, the rotation on the sky and the angular size were treated as free parameters \citep[for further details, see][]{2024A&A...692A.218M}. This exercise showed that a Roche-lobe-filling configuration provides a significantly better fit to the data, with reduced $\chi^2$ values ranging from 2.76 to 2.99. This improvement is particularly evident in the closure phases. Fits for models at orbital phases furthest from quadrature (30\textdegree) were less satisfactory, consistent with the expected near-symmetry of the projected stellar shape in those configurations. An illustrative model for a mass ratio of 2.4 observed at quadrature is shown in Fig.\ref{fig:ZZCMi}. If the giant in ZZ~CMi is indeed almost filling its Roche lobe, lower stellar masses would be favoured, and thus also a lower metallicity. A spectroscopic analysis of this object would be most useful.

\section{Conclusion}

We have measured the radius of the giant stars in seven symbiotic systems. Combined with the orbital elements, this has also allowed us to put these stars in the H-R diagramme. We derive several results from this (Tab.~\ref{tab:results}). First, except for ZZ CMi, all the giants are well within their nominal Roche lobe, even when taking into account the eccentricity of the orbit. Second, in most cases, the giant is already on the AGB, although there are a couple of cases where it is not possible to infirm that they are not on the tip of the RGB. In any cases, they are always quite evolved stars. Third, given their evolutionary stage and their orbital elements, even if they do not do so now, these systems will soon fill their Roche lobes and give way to double white dwarf systems, in either long or short orbits, depending on whether they will manage to escape a common envelope evolution. 
We finally confirm the results from \citetalias{2025A&A...695A..61M} that most of these giants are not synchronised. 

Our results depend crucially on the precise knowledge of their effective temperature, metallicity, and distances. For the former two, we encourage the community to provide more accurate and model-free atmospheric parameters for all known symbiotic systems. This will also allow us to better constrain stellar evolution models. 
Concerning the distance, it is our hope that \textit{Gaia} DR4 will, finally, deliver and be able to categorise these stars as binaries — not only spectroscopic, but also astrometric — thereby providing the orbital inclinations and thus the masses of the components, as well as a parallax corrected for the orbital motion. 
Even if for some reasons, \textit{Gaia} DR4 does not derive an orbit, the fact that we will have the individual measurements will allow us to reassess the parallax, taking into account the known orbital elements, and therefore be more accurate in our radius and luminosity estimates, and thus on the inferred masses. 

With the radii we derive here for the giant, one will then be able to obtain a final answer on the Roche lobe filling of these interesting objects, providing strong constraints on their past and future, and linking them with other families of binary stars.  

\begin{acknowledgements}
The research of J.M. was supported by the Czech Science Foundation (GACR) project no. 24-10608O and by the Spanish Ministry of Science and Innovation with the grant no. PID2023-146453NB-100 (PLAtoSOnG). We acknowledge the use of the Jean-Marie Mariotti Center \texttt{LITpro} service co-developed by CRAL, IPAG and LAGRANGE. This research has made use of the SIMBAD database, CDS, Strasbourg Astronomical Observatory, France. This work has made use of data from the European Space Agency (ESA) mission
{\it Gaia} (\url{https://www.cosmos.esa.int/gaia}), processed by the {\it Gaia}
Data Processing and Analysis Consortium (DPAC,
\url{https://www.cosmos.esa.int/web/gaia/dpac/consortium}). Funding for the DPAC
has been provided by national institutions, in particular the institutions
participating in the {\it Gaia} Multilateral Agreement.
\end{acknowledgements}

\bibliographystyle{aa} % style aa.bst
\bibliography{bibliography}

\end{document}